\documentclass[aps,amssymb,amsfonts,twocolumn]{revtex4}
\usepackage{amsmath}
\usepackage{graphicx}
\usepackage{amsfonts}
\usepackage{amssymb}

\def\textbf#1{{\bf #1}}
\def\be{\begin{equation}}
\def\ee{\end{equation}}
\def\ben{\begin{eqnarray}}
\def\een{\end{eqnarray}}
\def\eea{\end{array}}
\def\bea{\begin{array}}
\newcommand{\bei}{\begin{itemize}}
\newcommand{\eei}{\end{itemize}}

\begin{document}

\title{Towards efficient algorithm deciding separability
of distributed quantum states.}

\author{Piotr Badzi{\c{a}}g$^{1}$, Pawe{\l} Horodecki$^{2}$,
and Ryszard Horodecki$^{3}$} \affiliation{$^{1}$
Department of Mathematics and Physics, M\"{a}lardalens
H\"{o}gskola, S-721 23 V\"{a}ster\aa s, Sweden, \\
$^{2}$ Faculty of Applied Physics and Mathematics,
Technical University of Gda\'nsk, 80--952 Gda\'nsk, Poland, \\
$^{3}$ Institute of Theoretical Physics and Astrophysics,
University of Gda\'nsk, 80-952 Gda\'nsk, Poland}

\begin{abstract}
It is pointed out that separability problem for arbitrary
multi-partite states can be fully solved by a finite size,
elementary recursive algorithm. In the worse case scenario, the
underlying numerical procedure, may grow doubly exponentially with
the state's rank. Nevertheless, we argue that for generic states,
analysis of concurrence matrices essentially reduces the task of
solving separability problem in $m \times n$ dimensions to solving
a set of linear equations in about $\binom{mn+D-1}{D}$ variables,
where $D$ decreases with $mn$ and for large $mn$ it should not
exceed $4$. Moreover, the same method is also applicable to
multipartite states where it is at least equally efficient.
\end{abstract}


\maketitle
\section{Introduction}
Presence of quantum correlations in a distributed system is
probably the most clear and fundamental marker of the system's
non-classicality. In applications, quantum correlations make
grounds for the emerging field of quantum information technology.
In either context, be it fundamental or utilitarian, the notion of
quantum entanglement plays the key role \cite{Nielsen00}. Not
surprisingly then, a search for an unambiguous price-tag for the
entanglement present in a given arbitrary mixed distributed state
has been one of the important tasks of physics of information. It
appears, however, that this task is notoriously difficult even in
the seemingly simple case of bipartite states. A yet simpler
problem of a {\it finite, operational} necessary and sufficient
separability test for bipartite states in arbitrary dimensions is
still difficult.

The claim that an arbitrary bipartite $n \otimes m $ state
$\varrho$ \emph{is separable} means that this state can be
expressed as a convex combination of projectors on product
directions in the bipartite Hilbert space $\mathcal{H}_{AB}$. This
can be formally expressed as the following condition:

The statement
\begin{equation}\label{criter1}
\begin{array}{c}
  \exists (\psi_i^{\mu}, \phi_j^{\mu};\ i=1 \ldots n,j=1 \ldots m,\ \mu= 1 \ldots
  (mn)^2-1): \\
  \sum_{\mu=1}^{(mn)^2-1} \psi_i^{\mu}
  \phi_j^{\mu} (\psi_\alpha^{\mu}
  \phi_\beta^{\mu})^{\ast} = \varrho_{ij,\alpha \beta}
\end{array}
\end{equation}
is true.

To decide the truth value of statement (\ref{criter1}) is usually
difficult, nevertheless it is always possible in principle. The
problem to solve is an example of a \emph{decision problem for the
existential theory for reals} with $s=(mn)^2-1$ (number of real
parameters in $\varrho$) quartic (degree $d=4$) polynomials in
$k=2(m+n)(m^2n^2-1)$ real variables. As such, the problem
represents a special case of \emph{the quantifier elimination
problem}. The fact that there exist algorithms solving these
problems was first proved by A.~Tarski \cite{Tarski51}. Complexity
of his algorithm was, however, not elementary recursive. It means
that it could not be bounded from above by any power of
exponentials of a finite hight.

Although considerable progress in the area has been made over the
years and there are working algorithms for the quantifier
elimination problem implemented into such programs like MAPLE or
Mathematica, the existing algorithms are far too inefficient to
solve non-trivial separability problems. The best general purpose
algorithm which we have found is by Basu, Pollack and Roy
\cite{BPR96} and it solves the decision problem for the
existential theory using $s^{k+1}d^{O(k)}$ arithmetic operations.
For separability of a $3\times 3$ system formulated as in
statement (\ref{criter1}) this means an astronomical number of
operations, of the order of $81^{961}\cdot 4^{O(960)}\approx
10^{O(2412)}$.

The line of research aiming at a finite operational (in principle
analytic) necessary and sufficient criterion of entanglement in
arbitrary distributed states relies on the hope that the head on
attack on the problem described above is unnecessarily
complicated. Indeed, even for a pair of qubits, solving the
decision problem (\ref{criter1}) directly would require
$16^{121}\cdot 4^{O(120)}\approx 10^{O(220)}$ operations to
determine entanglement. One knows, however, that in this case as
well as for qubit-qutrit pairs, a simple partial transpose
criterion solves the problem efficiently \cite{MPR96}. Moreover,
in case of two qubits, Wootters found an efficient finite
algorithm to determine an optimal \emph{decomposition} of an
arbitrary 2-qubit state \cite{Wootters98}. These results suggest
some natural approaches to the separability problem in arbitrary
dimensions.

One may thus seek for a method which would reduce separability to
an eigenvalue problem, like in the partial transposition (PT) test
\cite{Peres}. Among the possible extensions and derivatives of PT,
the hierarchy of the PT tests \cite{Spedalieri02} as well as the
matrix realignment criterion \cite{CCN,CCN1} and its extensions
\cite{PartialCCN,PartialCCN1} have the desired, operational (ie.
eigenvalue-like) form. Moreover, these tests are generally more
powerful detectors of entanglement than PT. Nevertheless, they
still do not guarantee any universal necessary and sufficient
condition for separability. Entanglement witnesses method
\cite{Terhal,MPR96} and its optimizations \cite{Lewenstein}
(important form experimental point of view) have similar
drawbacks, although here, an exhaustive search over all the
possible pure separable states in principle solves the problem.

To avoid searching over an uncountable set, Gurvits reformulated
the separability problem as a weak membership problem
\cite{Gurvits} and showed that even in this limited guise, the
problem is still NP-hard. In the development which followed,
Ioannou at al. proved the existence of an, in principle, efficient
algorithm solving the underlying weak membership problem
\cite{I04}. The structure of the algorithm was, however,
complicated, which led to technical difficulties with its
implementation.

Hulpke and Bru\ss \ \cite{Hulpke04} showed that irrespective of
the required accuracy, the search may be limited to the states
with rational coordinates in a given basis. This result combined
with the symmetric extensions separability criterion by Doherty et
al. \cite{Spedalieri02} guaranteed a finite stopping time
(although no upper bound on it) for a separability test on all the
states but those exactly on the border between separable and
entangled.

Wootters' solution to the separability of two qubits relied on the
concept of concurrence and its generalization for mixed states,
concurrence matrix \cite{Wootters98}. In particular, introduction
of concurrence matrix allowed him to derive a simple closed
formula for entanglement of formation in terms of the singular
values of the matrix. After several attempts (see \cite{Rungta01})
the concept of concurrence matrix was extended to arbitrary
bipartite systems \cite{Badziag02}. There, it appeared that a
bilinear combination of concurrence matrices, a {\it biconcurrence
matrix} maybe the fundamental object to investigate in connection
with separability of bipartite quantum stats. This line of thought
was later confirmed by Mintert, Ku\'{s} and Buchleitner who used
biconcurrence to formulate powerful tests for entanglement of
bipartite systems \cite{Mintert04}. Recently, the same authors
successfully generalized biconcurrence to cover multipartite
states as well \cite{Mintert04b}. This led to powerful tests for
different kinds of multi-partite entanglement. However a tight
necessary and sufficient operational test still remained out of
reach, since it would usually need optimization procedures.

In what follows, we reexamine the role of biconcurrence in solving
separability problem for arbitrary distributed states. In
particular, we study the underlying decision problem formulated in
terms of the eigenvectors of biconcurrence matrix and show that
not only for low rank density matrices (like it was in the
analysis of range criterion \cite{PH00, PH97}), but also for
generic density matrices in many dimensions the resulting
solutions are computationally manageable.

In exceptional cases, the resulting numerical calculations may
still grow super exponentially with the increasing
$\textrm{rank}(\varrho)$. Therefore we do not claim a complete
solution to the separability problem. Nevertheless, we can quote
evidence that for generic multipartite states the calculations
grow polynomially with the size of the investigated system. An
improvement of the method which would entirely remove the
difficult exceptional cases can hardly be expected since
separability problem appears to be computationally hard \cite{Gurvits}.

\section{Bipartite separability and concurrence.}

Entanglement of pure bipartite states is well understood, so let
us begin by considering such a state

\begin{equation}\label{b-pure}
  |\psi \rangle_{AB} = \sum_{i,j} a_{ij}|i \rangle_{A} \otimes |j \rangle_{B}
\end{equation}

(the $i$'s and the {j}'s here number orthonormal basis vectors in
the Hilbert spaces held by Alice (A) and Bob (B) respectively).
Separability of $|\psi \rangle_{AB}$ means that the right hand
side of (\ref{b-pure}) factorizes, i.e., that
\begin{equation}\label{b-p-fact}
  |\psi \rangle_{AB} = \sum_{i,j} f_{i}|i \rangle_{A} \otimes g_{j}|j \rangle_{B}
\end{equation}
This is equivalent to the following conditions for the matrix
elements $a_{ij}$
\begin{equation}\label{b-p-f-1}
  \forall (i,j,k,l)\  a_{ij} a_{kl} - a_{il} a_{kj} = 0
\end{equation}
For a pure state in $m \times n$ dimensions, equation
(\ref{b-p-f-1}) imposes $\mathcal{N}=\frac{mn(m-1)(n-1)}{4}$
non-trivial conditions. They correspond to the requirement that
all the $2\times 2$ minors of matrix $[a]$ are zero. The degree to
which a given pure state violates condition (\ref{b-p-f-1}) can
thus be measured by the length of the $\mathcal{N}$-dimensional
vector $\mathbf{C}$ with components
\begin{equation}\label{Conc1}
 C_{\sigma} = C_{i\wedge k, j \wedge l} = 2(a_{ij} a_{kl} - a_{il} a_{kj})
\end{equation}
This length,
\begin{equation}\label{Conc2}
  C=\sqrt{\sum_{\sigma} |C_{\sigma}|^2}
\end{equation}
coincides with concurrence defined for pure multi-dimensional
systems by P.~Rungta et al. \cite{Rungta01}.

Separability of a mixed state $\varrho$ means that there exist a
decomposition of $\varrho$, where all the contributing pure states
have vanishing concurrence (\ref{Conc2}). One should notice here
that the decomposition of a given state is not unique. In fact,
any two ensembles of (sub-normalized) pure states $\{ |\psi^{\mu}
\rangle \}$ and $\{ |\phi^{\mu} \rangle \}$ realize the same state
$\varrho = \sum_{\mu} |\psi^{\mu} \rangle \langle \psi^{\mu} |$
\emph{iff} they are related by a unitary transformation
\cite{Schrod36}
\begin{equation}\label{DeCh}
  |\phi^{\mu} \rangle = \sum_{\nu} U_{\mu \nu} |\psi^{\mu} \rangle
\end{equation}

To decide existence of a separable decomposition of a given state
is in general difficult. One of the sources of this difficulty may
be associated with the fact that the set of pure state
concurrences contributing to a mixed state is not closed with
respect to the changes of the decomposition. The objects nearest
to $C$ which transform well with the decomposition changes are
concurrence matrices. For a given state decomposition
\begin{equation}\label{Decom1}
  \varrho = \sum_{\mu} |\psi^{\mu} \rangle \langle \psi^{\mu} |
\end{equation}
with sub-normalized pure components $|\psi^{\mu} \rangle =
\sum_{i,j} a_{ij}|i \rangle_{A} \otimes |j \rangle_{j}$, one can
define $\mathcal{N}$ symmetric concurrence matrices
\begin{equation}\label{ConMat}
  C_{\sigma}^{\mu\nu} = a_{ij}^{\mu} a_{kl}^{\nu} -
  a_{il}^{\mu} a_{kj}^{\nu} + a_{ij}^{\nu} a_{kl}^{\mu} - a_{il}^{\nu}
  a_{kj}^{\mu}
\end{equation}

When the state decomposition is changed from $\{ |\psi^{\mu}
\rangle \}$ to  $\{ |\phi^{\mu} \rangle \}$ via transformation
(\ref{DeCh}), then the concurrence matrices $C_{\sigma}^{\mu\nu}$
change into
\begin{equation}\label{ConMat2}
  \mathcal{C'}_{\sigma}^{\mu\nu} = \sum_{\alpha \beta} U_{\mu \alpha} C_{\sigma}^{\alpha
  \beta}U_{\mu \beta}
\end{equation}
So, in terms of concurrence matrices, separability means that
there exists such a state decomposition for which the diagonals of
all the $\mathcal{N}$ concurrence matrices $\mathbf{C}_{\sigma}$
are zero. One can immediately notice that in this context two
qubits represent an exceptional case: there is only one
concurrence matrix. This simplification allowed Wootters to
produce an elegant solution of the separability problem for two
qubits \cite{Wootters98}. Moreover, his solution resulted in a
simple and efficient algorithm for the determination of an optimal
decomposition of both separable and entangled two-qubit states.

The 2-qubits solution does not, however, generalize easily for
systems in higher dimensions. There, in terms of concurrence
matrices, a bipartite state is separable iff there exists isometry
$U$ such that
\begin{equation}\label{Sep2}
  C_{\sigma}'^{\mu \mu} = \sum_{\alpha \beta} U^{\mu \alpha} C_{\sigma}^{\alpha
  \beta} U^{\mu \beta} = 0
\end{equation}
for all values of $\sigma$ simultaneously.

It appears that the question of existence of $U$ which satisfies
the whole set of conditions (\ref{Sep2}) is much more difficult to
answer than that of existence of $U$ which satisfies only one of
these conditions (for a single given $\sigma$) solved in
\cite{Wootters98}. In particular, in a state with bound
entanglement, none of the $\mathbf{C}_{\sigma}$'s will show
entanglement separately, nevertheless, regardless of the choice of
the local bases, it will be impossible to find a decomposition
which zeros the diagonals of all the $\mathbf{C}_{\sigma}$'s
together.

To make the problem independent of the local choice of bases and
possibly reduce its complexity, one may consider a single object,
biconcurrence instead of the set of concurrence matrices
\cite{Badziag02}. For clarity, one may permute indexes in the
original definition (formulae (22) and (23) in \cite{Badziag02})
and define biconcurrence $\textbf{B}$ as

\begin{equation}\label{Bi-con}
  B^{\mu \nu;\alpha \beta} =\sum_{\sigma} C_{\sigma}^{\mu \nu} C_{\sigma}^{*\alpha
  \beta}
\end{equation}

(the asterisk denotes complex conjugate). Written in this form,
$\textbf{B}$ represents a manifestly positive operator in the
vector space of symmetric concurrence matrices.

In terms of biconcurrence, a state $\varrho$ is separable iff
given an initial decomposition of the state, e.g., the
eigendecomposition, there exists isometry $U$ such that
\begin{equation}
G_{0}=\sum_{i=1} \sum_{\mu \nu \alpha \beta=1}U^{i \mu} U^{i \nu}
B^{\mu \nu \ \alpha \beta} (U^{i \alpha} U^{i \beta})^{\ast} =0
\label{g0}
\end{equation}

This, together with the conditions making $U$ represent an
isometry, allows to associate entanglement with strict positivity
of a real quartic form in at most $2 mn ((mn)^2-1)$ variables. The
latter problem was formally solved by Jamiolkowski back in 1972
\cite{Jam72forms}. The level of complexity of this solution
was, however, to high to be of practical relevance in the
present context.

On the other hand, one may notice that existence of an isometry
$U$ solving (\ref{g0}) is equivalent to the existence of such $U$
which zeros the diagonals of all those eigenvectors of $B$
(hereafter denoted by $T_{\sigma}^{\mu\nu}$), which belong to
non-zero eigenvalues. By using biconcurrence in this way, Mintert,
Ku\'{s} and Buchleitner were able to derive simple, powerful lower
bounds on the entanglement of formation for arbitrary bipartite
states \cite{Mintert04}. In particular, it appeared that unlike
the $\mathbf{C}_{\sigma}$'s, single matrices $\mathbf{T}_{\sigma}$
can indicate bound entanglement, e.g., in the entangled states
introduced in \cite{PH97} and showed to be non-distillable in
\cite{BEnt}.

Concurrence matrices can be powerful indicators of entanglement
even when it is not seen in any single linear combination of
$\mathbf{T}_{\sigma}$ matrices, i.e., when the diagonal of every
linear combination of $\mathbf{T}_{\sigma}$'s can be brought to
zero separately. To see it, it is enough to notice that the
equation set (\ref{Sep2}) (or an equivalent equation set with the
$\mathbf{T}_{\sigma}$'s substituted for $\mathbf{C}_{\sigma}$'s)
represents a set of up to $\mathcal{N}$ homogeneous quadratic
equations for $r$ elements of the $\mu$'th row of isometry
$\mathbf{U}$ and that the elements $u_{\alpha}$ of each row
satisfy the same set of equations (when one begins with the
eigendecomposition then $r$ is equal to the state's rank).

\begin{equation}\label{EnCon1}
  \sum_{\alpha,\beta=1}^{r} T_{\sigma}^{\alpha
  \beta} u_{\alpha} u_{\beta} = 0, \ \ \sigma=1,2,..., \mathcal{N}
\end{equation}

When the solution set of this equation set is empty than one can
clearly claim entanglement right a way. Otherwise, whenever the
number of equations exceeds $r-2$, one may expect a finite number
of solutions, i.e., a finite number of possible rows of $U$
(modulo normalization). One should expect this to be the
prevailing situation since a generic state in $m \times n$
dimensions is of rank $r_g=mn$ while the corresponding number of
concurrence matrices is up to $\frac{mn (m -1) (n-1)}{4}$ which is
greater than $r_g$ for all the systems greater than two qubits, a
qubit and a qutrit and a qubit and a four-level system
(a clear indication that the bigger the space, the
more exceptional separable states are, as it was suggested
by numerical analysis  in \cite{Zycz}). To identify the variety of
solutions of the equation set (\ref{Sep2}) is the most costly
part of the procedure.

The necessary variable elimination involved in the process and
based on a standard construction of a Groebner basis is known to
be computationally difficult \cite{C-L-O'S} and even for
$0$-dimensional ideals, it may be polynomial in $d^n$, where $d$
is the maximum degree of the generators (original polynomials) and
$n$ is the number of variables \cite{Lak91}. This represents a
super-exponential growth and, like a brute force method indicated
in the introduction, it may lead to prohibitive calculations
already for small systems.

Fortunately, by employing the XL (eXtended Linearization) method
\cite{Courtois00}, one may drastically reduce the level of
complexity with growing $\Delta=\mathcal{N}-r+1$. The method
relies on the observation that the set (\ref{EnCon1}) can be
regarded as a set of $\mathcal{N}$ linear equations for
$\binom{r+1}{2}$ variables $x_{\alpha \beta} =
u_{\alpha}u_{\beta}$. Moreover, by multiplying each of the
equations by all the possible products $u_{1}^{p_1} \cdots
u_{r}^{p_r}$, such that $\sum_i p_i = D-2$, one can expand the
original set into a homogeneous set of
$\mathcal{N}\binom{r+D-3}{D-2}$ linear equations for
$\binom{r+D-1}{D}$ variables $x_{\alpha_1,\ldots \alpha_D} =
u_{\alpha_1} \cdots u_{\alpha_D}$. The task is now to make the
number of equations big enough to eliminate all but the last $D+1$
variables, e.g., $u_{1}^{D},\ u_{1}^{D-1}u_{2}, \ldots,\
u_{2}^{D}$ and put, e.g., $u_1=1$, so that one is left with a
single polynomial equation in one variable.

Numerical tests of the XL method reported in \cite{Courtois00}
indicate that for $\Delta=0$, the degree $D$ of the single
variable polynomial necessary to solve in order to obtain the
solution set of (\ref{EnCon1}) grows with $r$ like $2^{r-1}$. This
puts bipartite separability of generic states in $3 \times 3$ and
even in $4 \times 4$ dimensions within a reach of a simple PC.
Actually, it may be even simpler than this. In $3 \times 3$
dimensions, a generic $r$ is $9$ just like a generic
$\mathcal{N}$, thus giving $\Delta = 1$. Numerical tests reported
in \cite{Courtois00} indicate that in this case $D = r$ which is
clearly manageable even for relatively large $r$. Moreover,
according to the same source, for $\Delta = 2$ one has $D \approx
\sqrt{r}$ and for $\mathcal{N} = \epsilon r^2$, one should expect
$D \approx \lceil 1/\sqrt{\epsilon} \rceil$. This indicates that
for generic cases in many dimension ($\mathcal{N} \approx r^2/4$)
one should not expect the numeric difficulty of the problem to
grow substantially with the dimensions of the local Hilbert
spaces.

When equation set (\ref{EnCon1}) is solved then, to decide
existence of the required isometry, one should arrange the
possible rows $(u_{k 1}, u_{k 2}, u_{k 3}, \ldots)$ normalized by
suitable factors $\lambda_k$, so that the resulting matrix
$\textbf{U}$ is an isometry. For that, the factors have to
satisfy:
\begin{equation}\label{final3a}
  \sum_k |\lambda_k|^2 u_{k i}u_{k j}^{\ast} = \delta_{i j}
\end{equation}
 This is a simple set of linear equations for $|\lambda_k|^2$
and it can be solved (separability) or proved unsolvable
(entanglement) without any problem.

\section{Beyond bipartite states.}

Our strategy easily generalizes to multipartite states with
arbitrary number of parties sharing the state. One only has to
analyze eigenvectors of the sum of all the independent positive
matrices $\hat{\mathcal{A}}_{jk}^{lm}$ defined in
\cite{Mintert04b} instead of the eigenvectors of a biconcurrence
matrix $B^{\mu \nu \ \alpha \beta}$ discussed earlier. In general,
this will increase the number of concurrence matrices to consider,
thus making full separability less likely. Everything else goes
like for bipartite states. For instance, for a 3-qubit state,
there are three matrices $\textbf{\hat{\mathcal{A}}}$ associated
with the three types of possible bi-partite correlations referred
to in \cite{Mintert04b} as $c_1^{(3)}$, $c_2^{(3)}$ and
$c_3^{(3)}$. The maximum possible rank of each of the matrices is
three (the symmetric projector is three dimensional). In a generic
three-qubit state, one may then expect nine concurrence matrices
($\mathcal{N}=9$) and $r_g = 8$. To reduce the number of variables
to one in the original equation set (\ref{EnCon1}) one will then
need such $D$ that $\binom{r+D-1}{D}-D \leq \mathcal{N}
\binom{r+D-3}{D-2}$. Here $D=5$ satisfies the inequality. In
general, the bigger the system, the lower $D$ will be sufficient.
The difficult part of the algorithm is in the number of the
(linear) equation and variables to consider. In the example above,
one should expect them to be $1080$ and $792$ respectively.

\section{Conclusions.}

We have pointed out that separability problem for arbitrary
multi-partite states can be fully solved by a finite size,
elementary recursive algorithm. However, in the worst case
scenario, the underlying numerical procedure may grow
super-exponentially with the state's rank.

In an attempt to reduce this complexity, we investigated how
analysis of concurrence matrices of given bipartite and even
multipartite states may lead to exhaustive analytic separability
checks. It appeared that for generic states in $m \times n$
dimensions, analysis of concurrence essentially reduced the task
of solving separability problem to solving a set of linear
equations in about $\binom{mn+D-1}{D}$ variables, where $D$
decreases with $mn$ and for large $mn$ it should not exceed $4$.
Moreover, the the same method is also applicable to multipartite
states where it is at least equally efficient.

One can notice, however, that the relatively low expected
complexity of the analysis of the generic states may be nothing
more than a reflection of the fact that these states are usually
entangled. Indeed, when the variety of the solutions of
(\ref{EnCon1}) is either zero-dimensional or empty, then there is
not much room for (\ref{final3a}) to have any solutions. One
possibility to extend the present results can then aim at gaining
some understanding of the intermediate case when there are many
independent concurrence matrices, nevertheless the variety of
solutions of (\ref{EnCon1}) is at least one-dimensional. It would
also be interesting to see whether in the language of concurrence
it is possible to perform decomposition of the state into a
separable part and a low dimensional part like it was for the edge
states in optimization process \cite{Lewenstein}. A possibility
like this would realize a version of Lewenstein-Sanpera
decomposition \cite{LS98}.

We thank M. Horodecki and K. Horodecki for fruitful
discussions. The work was supported by EU project RESQ
(No. IST-2001-37559) and by Polish Ministry of Scientific
Research and Information Technology under the (solicited)
project No. PBZ-MIN-008/P03/2003.

\end{document}